\documentclass[aps,prl,twocolumn,preprintnumbers,showpacs]{revtex4}

\usepackage{psfrag,graphicx}
\usepackage{dcolumn}
\usepackage{amsmath,amssymb}
\usepackage{bm}
\usepackage{amsfonts,amssymb,amsmath}        

\newcommand{\be}{\begin{equation}}
\newcommand{\ee}{\end{equation}}
\newcommand{\bq}{\begin{eqnarray}}
\newcommand{\eq}{\end{eqnarray}}
\newcommand{\I}{\mathbb {I}}

\newcommand{\e}{\text{e}}

\newcommand{\ket}[1]{\left | \, #1 \right\rangle}
\newcommand{\bra}[1]{\left \langle #1 \, \right |}

\newcommand{\proj}[1]{\ket{#1}\bra{#1}}
\newcommand{\Sp}{\,\,\,\,\,\,}
\newcommand{\no}{\nonumber\\}

\newtheorem{lemma}{Lemma}

\def\qed{\leavevmode\unskip\penalty9999 \hbox{}\nobreak\hfill
     \quad\hbox{\leavevmode  \hbox to.77778em{%
               \hfil\vrule   \vbox to.675em%
               {\hrule width.6em\vfil\hrule}\vrule\hfil}}
     \par\vskip3pt}

\begin{document}

\title{Stochastic Matrix Product States}

\author{Kristan Temme}
\author{Frank Verstraete}
\affiliation{Faculty of Physics, University of Vienna, Boltzmanngasse 5, A-1090 Vienna, Austria}
\date{\today}

\begin{abstract}
The concept of stochastic matrix product states  is introduced and a natural form for the 
states is derived. This allows to define the analogue of Schmidt coefficients for steady states 
of non-equilibrium stochastic processes.  We discuss a new measure for correlations 
which is analogous to the entanglement entropy, the entropy cost $S_C$, and show that this measure 
quantifies the bond dimension needed to represent a steady state as a matrix product state. 
We illustrate these concepts on the hand of the asymmetric exclusion process.
\end{abstract}
\maketitle

This paper explores parallels between the many-body description of non-equilibrium  
steady states of classical stochastic processes and ground states of strongly correlated 
quantum many-body systems. Classical non-equilibrium  steady states are typically  
much richer than their equilibrium counterparts and can exhibit interesting  behavior 
such as the presence of a current, non-equilibrium phase transitions and entire phases 
with a diverging correlation length \cite{ASEP_MF,ASEP2}.
There are strong parallels between the many-body description of  non-equilibrium classical 
stochastic spin systems \cite{Felderhof}, and it turns out that matrix product states 
(MPS) play a very important role in both fields \cite{EvansMPS,Adv}.

The theory of entanglement,  developed in the context of
quantum information theory, has recently proven to yield valuable
new insights in the nature of the wavefunctions arising in such strongly
correlated quantum many-body systems. One of its main virtues is
that it allows to quantify the amount of quantum correlations in a many-body system
in terms of so--called area laws  \cite{VidalArea,Cardy,areaRMP} and relate this to the 
structure of the underlying wavefunctions representing the ground states of quantum spin Hamiltonians 
\cite{FrankFaith,Hastings}. The ground states of quantum spin chains is well described by the class of MPS, 
since for local Hamiltonians the associated Schmidt coefficients of the ground state decay very fast. 
A MPS approximation with finite bond dimension can be obtained by setting the smallest Schmidt coefficients 
$\sigma_I$ equal to zero. This approximation is justified whenever an area law is satisfied \cite{FrankFaith}. 
If we define $p_i=\sigma_i^2$, then those $p_i$ form a probability distribution. The entropy of this
distribution quantifies the dimension of the matrices that has to be chosen in the MPS approximation 
such as to obtain a given fidelity.  For such pure states, this entropy has multiple operational meanings,
ranging from the amount of distillable Bell states by local operations to the maximum amount of secret 
information that can be sent in a cryptographic setting \cite{Winter}.  For mixed states and in the presence 
of classical correlations the situation is more complicated. A lot of research in quantum information has concentrated 
on resolving the relationships between classical and quantum correlations in mixed quantum states. 
A measure  of particular importance to our approach is the entanglement of purification \cite{BTerhal}. 
It quantifies both quantum and classical correlations in terms of the number of maximally entangled 
states needed to asymptotically generate the state.

The main topic of this paper is to explore the classical non-equilibrium analogous of the quantum notions of entanglement
entropy, area laws and the density matrix renormalization group which justify the use of MPS in the 
quantum setting. The main technical difficulty arises from the fact that classical probability distributions are normalized in the $L_1$ 
norm ($\sum_i|p_i|=1$), while quantum states are normalized in the $L_2$ norm ($\sum_i|\psi_i|^2=1$). 
This  can partly be overcome by working with a subclass of MPS where all matrices only contain non-negative entries; 
we call this subclass stochastic matrix product states (sMPS). The concept of mutual information, defined for classical bipartite distributions
$p_{AB}$ where $A$ and $B$ will represent the variables or spins on both halves of a chain, 
plays a role analogous to the entanglement entropy: $I(A:B) = \sum_{AB} p_{AB}\log_2\left(\frac{p_{AB}}{p_{A} p_{B}}\right)$.
It immediately gives an upper bound to the error made when approximating $p_{AB}$ by a product of its marginals, since 
$\| p_{AB}-p_{A}p_{B} \|_1^2 \leq 2 \ln(2) I(A:B)$ \cite{CT}. Just as in the quantum case, one would expect that the 
global non-equilibrium steady state probability distribution of the stochastic process can be represented as a stochastic matrix 
product state (sMPS) with small bond dimension if this mutual information is small. However,  more subtle measures are 
needed in the case of stochastic processes, and we will introduce the notion of entropy cost
to quantify the bond dimension needed for the corresponding sMPS. 
\begin{paragraph}{{\bf Definition 1}
 The entropy cost $S_C$ for a bipartite probability distribution $P(x,y)$ is given by:}
 \be
 	S_C = \min_{p_\lambda,P_A,P_B} S(\{p_\lambda\})
\ee
{\it where $S(\{p_\lambda\}) = -\sum_{\lambda}p_{\lambda}\log_2\left(p_{\lambda}\right)$ is
the Shannon information of $\{p_\lambda\}$, and where the optimization is over all probability distributions 
$p_\lambda$ and over all conditional probabilities  $P_A$ and $P_B$ for which 
$P(x,y) = \sum_\lambda P_A(x | \lambda)P_B(y | \lambda)p_\lambda$.}\\
\end{paragraph}
The entropy cost bears a lot of resemblances  to the notion of common information introduced by Wyner in the context
of cryptography and classical information theory \cite{Wyner}, and this entropy cost serves as an upper bound to the common information.
The entropy cost can be thought of as the classical analogue of the entanglement of purification, and the probability distribution 
$p_\lambda$ plays a role analogous to the Schmidt coefficients in the quantum case.

To give an example, consider the equilibrium system defined by the classical Ising model  
$H = \sum_{i=1}^{N-1}s_i s_{i+1}$,  with $s_{i} = \pm 1$. The equilibrium distribution $p(\{s_i\})_{I} =  \exp(-\beta H)/Z$
can be written in terms of a very simple MPS with  $D=2$ \cite{FrankPEPS}; the entropy cost $S_{C}$ can then easily be 
calculated exactly in the case $N=2$ and is given by 
\bq
\label{SCIsing}
S_{C} = -\left(\e^{-\beta}\cosh(\beta)\log_2\left(\e^{-\beta}\cosh(\beta)\right) + \right.\no
\left. \e^{-\beta}\sinh(\beta)\log_2\left(\e^{-\beta}\sinh(\beta)\right)\right).
\eq 
As expected, this function monotonously increases from $0$ to $1$, i.e. from the paramagnetic without 
correlations to the ferromagnetic region with strong correlations. A different measure that was recently 
investigated is the shared information \cite{shared}. It has been shown that it obeys an area law for 
several non-critical stochastic models and that critical behavior can be identified by logarithmic corrections.

Let us next define  a $D$-dimensional sMPS describing a 
classical probability distribution of $N$-spins each of dimension $d$; obviously, those sMPS 
were already extensively used in the literature, and we are just formalizing the definition here.
\begin{paragraph}{{\bf Definition 2}
 A stochastic matrix product state (sMPS) is given by:}
\be
\label{MPSgen}
\ket{p_D} = \sum_{i_1,\ldots,i_N=1}^{d} \bra{L} B^{1}_{i_1} \ldots B^{N}_{i_N} \ket{R} \ket{i_1 \ldots i_N},
\ee
{\it where we only consider real matrices that are $D_k\times D_{k+1}$ dimensional, 
with $D_k \leq D$, and additionally fulfill the requirement 
$[B^{k}_{i_k}]^{\gamma}_{\delta} \geq 0$ for every element individually.}
This ensures that all the weights of the distribution 
are positive after contraction. The left and right vector $\bra{L}$ and $\ket{R}$ are 
also elementwise positive and can be absorbed into the matrices $B^1_{i_1}$ and $B^N_{i_N}$, 
which corresponds to choosing $D_1=D_N =1$. Furthermore, we require $\ket{p_D}$ to be 
normalized in the  $L_1$ norm, $\| \ket{p_D} \|_1 = 1$.\\
\end{paragraph}
Every multipartite probability distribution of a chain 
of discrete variables can obviously be written in the form (\ref{MPSgen}) if we allow for 
a sufficiently large matrix dimension $D^{\max}$, i.e. exponential in the number of sites. 

\begin{center}
\begin{figure}[ht]
\vspace{-0.8cm}
\resizebox{1.05\linewidth}{!}
{\includegraphics{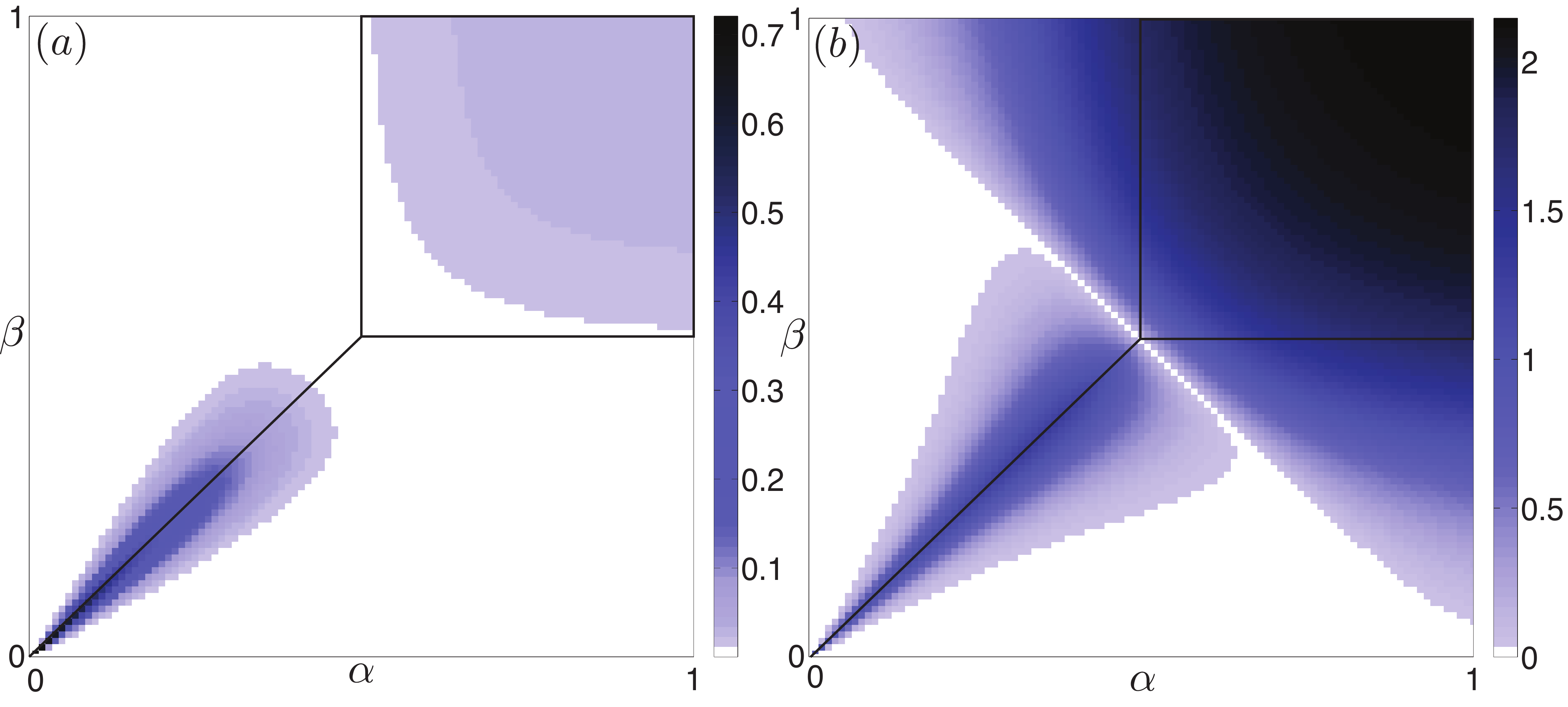}}
\caption{\label{phase_map}(color online) Figure (a) The mutual information of a chain 
length $N=20$ for different inflow parameters $\alpha$ and $\beta$ has been calculated numerically. 
Figure (b) entropy cost for a representation with $D^{\max}=21$. Note the different scales of the two plots.
The resulting plots clearly reflect the underlying phase diagram of the ASEP \cite{Derrida},
both for the mutual information and the approximate entropy cost. 
The entropy cost  as well as the actual mutual information drop to zero along the mean-field line $\alpha+\beta=1$. 
The mutual information is consistently low throughout the diagram, explaining 
why mean-field approaches have given such good results \cite{ASEP_MF}.}
\vspace{-1cm}
\end{figure}
\end{center}

Let us next show how the entropy cost can be calculated. Upon inserting a partition of 
unity $ \I = \sum_{\lambda=1}^{D_k} \proj{\lambda}$ in  (\ref{MPSgen}), 
we can write $\ket{p} = \sum_{\lambda}\sum_{\{i_n\}}\bra{L} B^{1}_{i_1} 
\ldots B^{k}_{i_k}\proj{\lambda}B^{k+1}_{i_{k+1}} \ldots B^{N}_{i_N}\ket{R} 
\ket{\{i_n\}}$. Now \vspace{-0.5cm}
\be
\label{singVal}
p_{\lambda} = \bra{L}\prod_{n=1}^{k} C^{[n]}\proj{\lambda}\prod_{n=k+1}^{N}C^{[n]}\ket{R}.
\ee
defines a new probability distribution if we chose $C^{[n]} = \sum_{i_n}B^{[n]}_{i_n}$
(i.e. the transfer matrix). This allows us to rewrite the MPS as
\be\label{source}
\ket{p}=\sum_{\lambda=1}^{D_k}\sum_{i_1 \ldots i_N} P_A(\{i_n\}_{n \in A} | \lambda) p_{\lambda}
P_B(\{i_n\}_{n \in B}| \lambda)
\ee
where \vspace{-0.5cm}
\bq 
P_A(\{i_n\}_{n \in A} | \lambda) 
&=&\frac{\bra{L}B^{1}_{i_1}\ldots B^{k}_{i_k}\ket{\lambda}}{\bra{L}\prod_{l=1}^{k}C^{[l]}\ket{\lambda}}\no  
P_B(\{i_n\}_{n \in B}| \lambda) 
&=&\frac{\bra{\lambda}B^{k+1}_{i_{k+1}}\ldots B^{N}_{i_N}\ket{R}}{\bra{\lambda}\prod_{l=k+1}^{N}C^{[l]}\ket{R}}.
\eq
The probability distribution $\{p_{\lambda}\}$ sums up to one due to 
the normalization of (\ref{MPSgen}), and $P_A(\{i_n\}|\lambda)$ and $P_B(\{i_n\}|\lambda)$ can 
be interpreted as information channels. Note that there are several partitions 
of unity that will give rise to valid $P_A$,$P_B$ and $p_\lambda$, and therefore the 
decomposition (\ref{source}) is not unique. 
It is probably a NP-hard problem to find the optimal decomposition that minimizes the entropy cost, 
and in practice we will therefore rely on the construction that was just given for finding upper bounds to it. 
It is also easy to find lower bounds to the entropy cost:
\begin{lemma}
\label{EntroCost}
For a given distribution $\ket{p}$ the mutual information $I(A:B)$ 
is bounded by the entropy cost $S_C$:  $I(A:B) \leq S_C$.
\end{lemma}
{\emph Proof:}
By virtue of (\ref{source}) we can focus on calculating the mutual information $I(\lambda:\mu)$ 
of the distribution $P(\lambda,\mu) = p_{\lambda}\delta_{\lambda,\mu}$, since the full 
distribution can be read as $P_{AB} = \sum_{\lambda,\mu} P_A(\{i_{A}\}|\lambda)
P_B(\{i_{B}\}|\mu)P(\lambda,\mu)$. This corresponds to a source that generates two 
outputs, which are then transformed by the channels $P_A$ and $P_B$. The
data-processing inequality \cite{CT} guarantees that the mutual information of 
the processed source $I(A:B) = I(p_A(P):p_B(P)) \leq I(\lambda : \mu)$ is smaller than 
the mutual information of the source itself which is equal to its entropy $I(\lambda:\mu) = S(\{p_{\lambda}\})$. 
The decomposition $[P_A(\{i_A\}|\lambda),P_B(\{i_B\}|\lambda),p(\lambda)]$ is not unique, 
therefore the bound is improved by taking the minimum over all decompositions. \qed
The decomposition of the sMPS as given in (\ref{source}) suggests the existence of the 
following (non-unique) normal form:
\be
\label{natural}
\ket{p} = \sum_{i_1\ldots i_N} A^{[1]}_{i_1}P^{[1]}A^{[2]}_{i_2}\ldots P^{[N-1]}A^{[N]}_{i_N}\ket{i_1\ldots i_N}.
\ee
Here the matrices $P^{[k]}$ represent diagonal matrices with probabilities $\{p_{\lambda_k}\}$ sorted 
in decreasing order. The matrices $C^{[k]} = \sum_{i_k} A^{[k]}$, 
where $[A^{[k]}_{i_k}]^{\gamma}_{\delta} \geq 0$, combined with the $P^{[k]}$'s  
behave as stochastic matrices. That is, $P^{k-1}C^{[k]} = S^k$ and $C^{k}P^k = S_{k}^{T}$, 
where $S^k$ and $S_{k}$ denote different stochastic matrices.\\
To see that every MPS distribution can be written this way consider the following scenario: We start 
by introducing the first bipartitioning between the first two sites. After the necessary 
normalization we proceed to the next site and perform the same procedure renormalizing 
the resulting matrices by the total contraction of the two halves of the chain. Proceeding along  
the chain results in the desired form. 
\begin{center}
\begin{figure}[ht]
\resizebox{0.95\linewidth}{!}
{\includegraphics{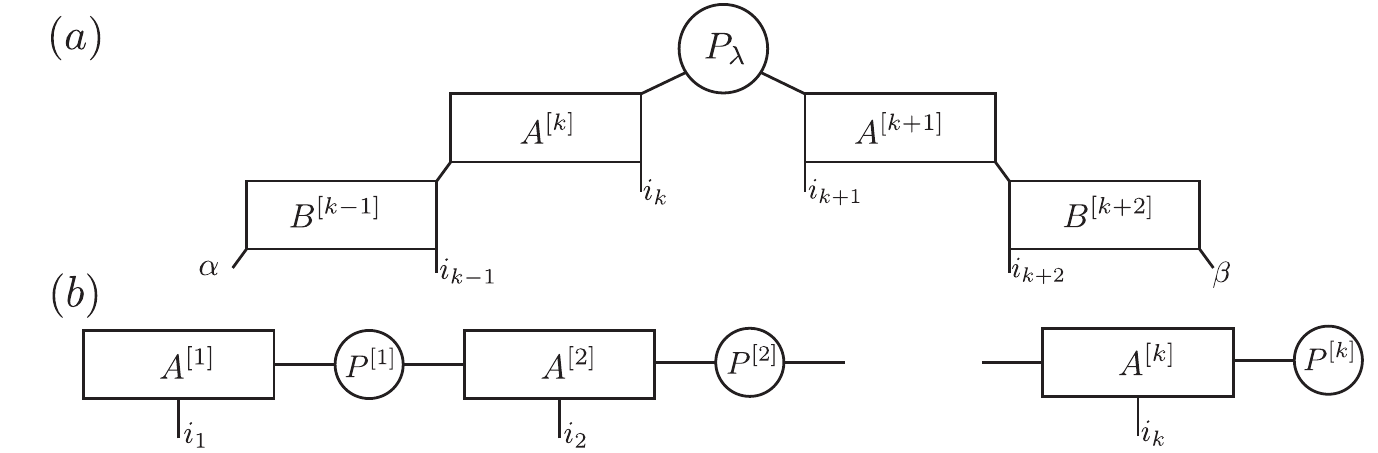}}
\caption{\label{fig:Channel}Pictorial representation of the natural sMPS decomposition:
Image (a) can be seen as the graphical representation of eqn. (\ref{source}). 
From a given source $P_\lambda$, the correlations are distributed via the two 
channels on the left and right. The normalizing factor is included in the $A$'s.
In (b) the analogy to the quantum MPS becomes evident for the decomposition as given in 
eqn. (\ref{natural}).The probabilities in the matrices $P^{[k]}$ are  the analogous 
of the singular values which arise upon a Schmidt decomposition of the quantum 
state \cite{Vidal}.}
\vspace{-1cm}
\end{figure}
\end{center}
This representation (\ref{natural}) 
enables us to give a good estimate on the error measured in the  
$L_1$ norm which is made upon truncating the dimension of the 
source space, i.e. neglecting probabilities smaller than a given 
value along each bipartition: 
\begin{lemma}
\label{LemErr}
For every multipartite distribution $\ket{p}$ there exists a MPS $\ket{p_D}$ 
of the form (\ref{natural}) with dimension $D$, such that \vspace{-0.4cm}
\[\left\| \ket{p} - \ket{p_D} \right\|_1 \leq 2 \sum_{k=1}^{N-1} \epsilon_k(D),\] 
where  $\epsilon_k(D) = \sum_{\lambda=D+1}^{D^{\mbox{max}}_{k}} p^{[k]}_{\lambda}$. 
\end{lemma} 
{\emph Proof:}
We can always write $\ket{p}$ as a distribution of the form (\ref{natural}) with a 
$D^{\mbox{max}}_{k} = d^{N}$. We now introduce another MPS $\ket{p_D}$
in natural form with a bond dimension of $D$. Let $\ket{p_D} = \ket{p^{*}_D}/\|\ket{p^{*}_D}\|_1$, 
where $\ket{p^{*}_D}$ is the unnormalized probability distribution which arises 
from neglecting along each cut all the probabilities $\{p_{\lambda_k}\}_{D+1}^{D^{\max}_k}$. 
We write $\ket{p^{*}_D} = \sum_{\{i_k\}} A^{[1]}_{i_1}P^{*[1]}\ldots A^{[N]}_{i_N}
\ket{\{i_k\}_{k=1}^N}$. Note, that if $\|\ket{p}-\ket{p^{*}_D}\|_1 \leq \epsilon$, then 
$\|\ket{p}-\ket{p_D}\|_1 \leq 2 \epsilon$ due to the triangle inequality.
Since $\ket{p^{*}_D}$ arose by only neglecting positive numbers, we may write 
$ \|\ket{p}-\ket{p^*_D}\|_1 = C^{[1]}P^1\ldots C^{[N]}-C^{[1]}P^{*1} \ldots C^{[N]}
= (\bra{l^{N-1}}-\bra{l^{N-1}}^*)C^{[N]} = \|(\bra{l^{N-1}}-\bra{l^{N-1}}^*)\|_1.$
Here we have used the fact that all summands are positive and we defined 
$\bra{l^{k}} = C^{[1]}P^{[1]}\ldots P^{[k]}$ as well as $\bra{l^{k}}^* = C^{[1]}P^{*[1]}\ldots P^{*[k]}$.  
The difference $\|\bra{l^1}-\bra{l^1}^{*}\|_1 = \|\bra{1\ldots 1}\left(P^{[1]}-P^{*[1]}\right)\|_1 = 
\sum_{\alpha_1 = D+1}^{D^{\max}_1}p_\lambda$ is simply given by $\epsilon_1(D)$. 
Note that due to (\ref{natural}) $\bra{l^{k-1}}C^{[k]} = \bra{1\ldots 1}$.
Proceeding to calculate the difference for other $k$ we find that
$ \left\|\bra{l^{[k]}}-\bra{l^{*[k]}}\right\|_1 \leq \left\|\left(\bra{l^{[k-1]}}
-\bra{l^{*[k-1]}}\right)C^{[k]}P^{[k]}\right\|_1 +\left\|\bra{l^{*[k-1]}}C^{[k]}\left(P^{[k]}-P^{*[k]}\right)\right\|_1 
\leq \sum_{n=1}^{k-1} \epsilon_n(D) +  \left\|\bra{1\ldots 1}\left(P^{[k]}-P^{*[k]}\right)\right\|_1.$
The last summand corresponds exactly to $\sum_{\alpha_k=D+1}^{D^{\max}_k} p_{\alpha_k} 
=\epsilon_k(D)$, which completes the proof. \qed
We have therefore proven that there exists an efficient parameterization of the 
steady state in terms of a sMPS with low bond dimension if there exists a parameterization of 
this steady state for which the entropy cost with respect to all bipartite cuts is small: 
if this is the case, then $\sum_{k=1}^{N-1}\epsilon_k(D)$ can be made small by following 
the arguments outlined in  \cite{FrankFaith}.  This is the analogue of the quantum case for 
which the existence of an area law implies the existence of an efficient representation in terms of MPS. 
Note however that the classical statement is a bit weaker, as the same normal form has to be used 
with respect to all bipartite cuts, and there is no guarantee that the same parameterization is optimal 
for all of bipartitions.\\
To make the investigations concrete, we consider the  non-equilibrium steady state of 
the asymmetric exclusion process (ASEP) \cite{Derrida}. This process is modelled by a chain of sites 
labeled $k = 1\ldots N$ occupiable by hardcore particles, i.e. classical spins $i_k \in \{0,1\}$. 
The particles are only allowed to hop to the right, and this only if the next site is empty. 
To drive the system,  particles at the left are injected with a given rate $\alpha$ and removed on the right with a 
rate $\beta$. This process exhibits three phases determined by the inflow $\alpha$ and outflow 
$\beta$. As was shown in \cite{Derrida} the steady state $\ket{p}$ of the corresponding 
master equation can be found exactly in terms of a MPS, albeit one for which the matrices 
are infinite dimensional and not necessarily positive.\\
The unnormalized steady state solution of the ASEP master equation \cite{Derrida} is given in MPS form (\ref{MPSgen}) 
with site independent matrices $B^1_{0} = \ldots = B^{N}_{0} = {\cal E}$ and $B^1_{1} = \ldots = B^{N}_{1} = {\cal D}$  
and boundary vectors $\bra{L},\ket{R}$, where all those satisfy the algebraic constraints ${\cal DE = E+D}$, 
$\bra{L}{\cal E} = \frac{1}{\alpha}\bra{L}$ and ${\cal D} \ket{R} = \frac{1}{\beta}\ket{R}$. 
Except for the special case when $\alpha+\beta = 1$, the representations obeying those constraints are infinite dimensional.
As the total occupation number of particles is limited by $N$, we will construct truncated representations of a given dimension 
$D = N+1$ that still reproduce the exact solution for a chain of length $N$; the entropy cost can then immediately be upper 
bounded by the logarithm of the system size, just as in the case of
 critical quantum spin chains. These representations are of  the form 
${\cal E} =\sum_{i,j=0}^{1}\left[B\right]_{i,j}\ket{i}\bra{j} + \sum_{n=2}^{N}\proj{n} + \ket{n}\bra{n-1}$ 
 and ${\cal D} =\sum_{i,j=0}^{1}\left[A\right]_{i,j}\ket{i}\bra{j} + \sum_{n=2}^{N}\proj{n} + \ket{n-1}\bra{n}$. 
 With the left and right vector $\bra{L} = \sum_{n=0}^1 w_n\bra{n}$ and $\ket{R}  = \sum_{n=0}^1 v_n\ket{n}$.
Here  $A$ and $B$  are 2-dimensional matrices fulfilling $AB + \sigma^{-}\sigma^{+} = A+B$ 
(here $\sigma^{+}$ and $\sigma^{-}$ are the Pauli raising and lowering operators),
and the 2-component vectors $v$ and $w$ must be chosen to be eigenvectors:  
$Av = (1/\beta) v$,  $wB = (1/\alpha) w$.\\
The construction given in  (\ref{source})  gives us a way of bounding the entropy cost as a 
function of the parameters $\alpha,\beta$ and $N$. We varied over all $A,B,v,w$ satisfying 
the algebraic relations, and found that the minimum is obtained for 3 different solutions 
depending on the parameter range of  $\alpha$ and $\beta$. The solution is 
depicted in Fig \ref{phase_map}.  
If $\alpha+\beta \leq 1$ and $\beta \leq \alpha$, then $A=1/\beta \proj{0} +\proj{1}$, 
$B = 1/(1-\beta) \proj{0} + b \ket{1}\bra{0} + 1/\alpha \proj{1}$, $w =(\alpha(1-\beta)\Sp b)$ and
$v^T = (1 \Sp 0)$ with $b=\sqrt{1-\alpha-\beta}$. However, if $\alpha+\beta \leq 1$ but 
$\beta \geq \alpha$, the optimal solution can be obtained from the previous one by the 
replacements, $\tilde{A} = B^T(\alpha \rightleftarrows \beta)$, 
$\tilde{B} = A^T(\alpha \rightleftarrows \beta)$, and $\tilde{w} = v^T$, 
$\tilde{v}=w^T(\alpha \rightleftarrows \beta)$. 
For $\alpha+\beta \geq 1$,  the optimal solution is given by
$A=1/\beta \proj{0} + a\ket{0}\bra{1} +\proj{1}$, $B=1/\alpha \proj{0} + a\ket{1}\bra{0} +\proj{1}$, 
$v^T = w = (1 \Sp 0)$ and $a = \sqrt{1/\alpha + 1/\beta -1/\alpha\beta}$.\\
\begin{center}
\begin{figure}[ht]
\resizebox{\linewidth}{!}
{\includegraphics{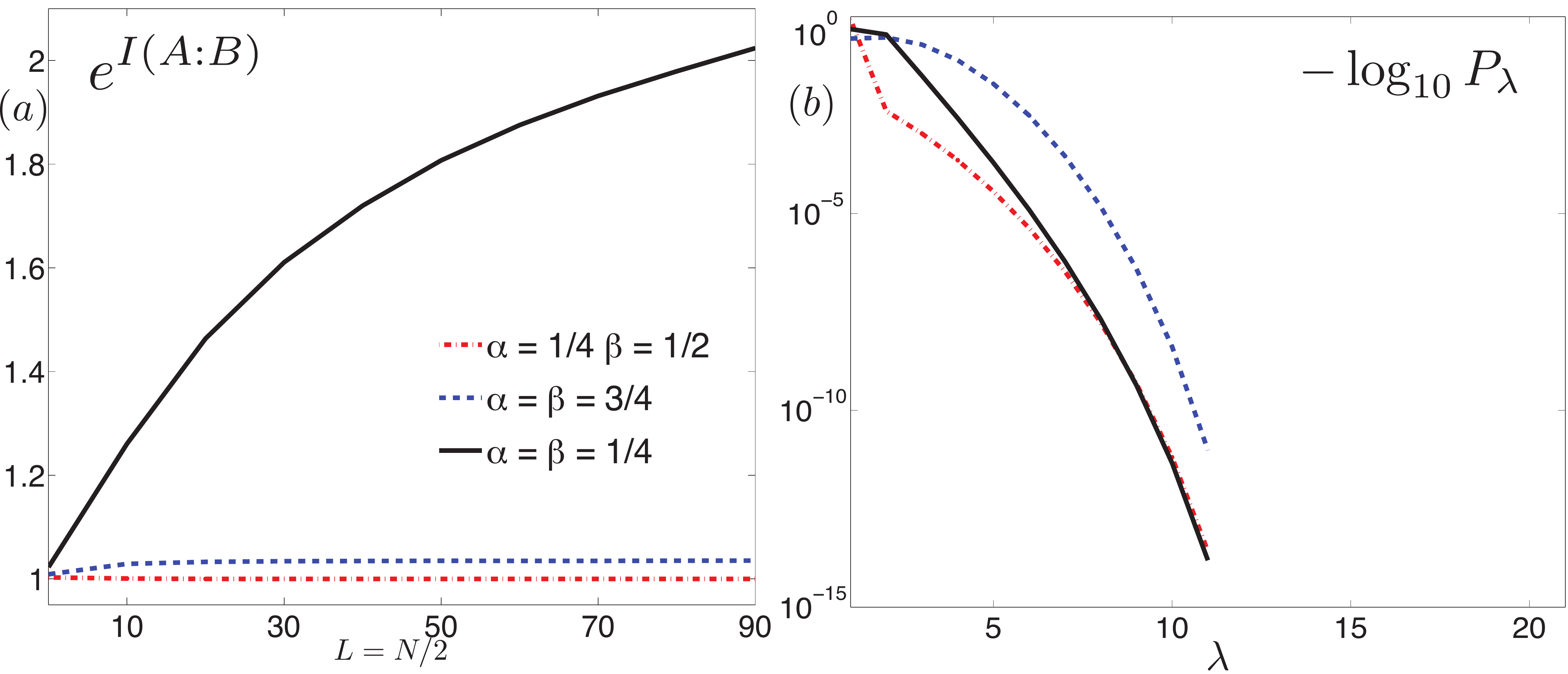}}
\caption{\label{fig:probScale} (color online) Plot (a): Monte Carlo simulations of the mutual information for different  $\alpha$
and $\beta$ values. The system size was varied from $N = 2\ldots 180$. The chain was cut in the middle, 
the length $L$ of the individual blocks are given by half of the total system size. The plot shows the exponential of 
the mutual information $\exp{I(A:B)}$.  The simulations  suggest that the mutual information grows only sub-logarithmically 
along the coexistence line $\alpha=\beta$, $\alpha+\beta \leq 1$;  Plot (b): The logarithm of the Probability distribution $\{p_\lambda\}$
plotted for different values of $\alpha$ and $\beta$. The dimension of the matrices $\cal E$ and $\cal D$
is $D^{\max} = 21$, corresponding to a chain of length $N=20$. The distributions decay super-exponentially. For all 
$\lambda \geq 11$, $p_{\lambda} = 0$.} 
\vspace{-0.6cm}
\end{figure}
\end{center}
\vspace{-1cm}
{\em Conclusions:} We have revisited the notion of stochastic matrix product states, and showed that a 
low bond dimension suffices to efficiently parameterize steady states of non-equilibrium distribution 
if the entropy cost in the system is low. This opens up the interesting question to characterize the 
conditions under which such steady  states have a low entropy cost. It would be interesting to see 
to what extent this relates to the gap of the corresponding stochastic process. This also opens up novel 
ways for constructing numerical renormalization group methods for simulating non-equilibrium systems 
in the line of the MPS algorithms for quantum spin chains \cite{White,schollwoeck,Adv}.

{\em Acknowledgements:}
The authors thank J.I. Latorre, Renato Renner and Andreas Winter for valuable discussions. 
This work was supported by the FWF doctoral program Complex Quantum Systems (W1210), 
the FWF SFB project FoQuS, and the European STREP grant Quevadis and ERC grant QUERG. 

\end{document}